\documentclass{article}
\usepackage{CJKutf8}   
\usepackage{spconf,amsmath,graphicx,hyperref}
\usepackage{amsfonts}

\usepackage{boldline} 

\usepackage{cuted} 
\usepackage{tabularx}

\title{Deep Dubbing: End-to-End Auto-Audiobook System with Text-to-Timbre and Context-Aware Instruct-TTS}
\name{%
  \parbox{\linewidth}{%
    \centering
    \itshape
    Ziqi Dai$^{1\dagger}$ \quad
    Yiting Chen$^{2\dagger}$ \quad
    Jiacheng Xu$^{2}$ \qquad
    Liufei Xie$^{2}$ \qquad
    Yuchen Wang$^{2}$ \qquad
    Zhenchuan Yang$^{2}$ \qquad
    Bingsong Bai$^{3}$ \qquad
    Yangsheng Gao$^{2}$ \qquad
    Wenjiang Zhou$^{2}$ \qquad
    Weifeng Zhao$^{2\star}$ \qquad
    Ruohua Zhou$^{1\star}$ \qquad
  }%
}
\address{
  \vspace{-2em}
  \thanks{${\dagger}$Equal contribution $^{\star}$Corresponding authors. } \\ 
  $^{1}$Beijing University of Civil Engineering and Architecture, Beijing, China \\
  $^{2}$Tencent Music Entertainment Lyra Lab, Shenzhen, China \\
  $^{3}$Beijing University of Posts and Telecommunications, Beijing, China \\
}

\usepackage{afterpage} 

\usepackage{multicol}

\begin{document}
\begin{CJK}{UTF8}{gbsn}
%

\ninept

\maketitle
\begin{abstract}

The pipeline for multi-participant audiobook production primarily consists of three stages: script analysis, character voice timbre selection, and speech synthesis. Among these, script analysis can be automated with high accuracy using NLP models, whereas character voice timbre selection still relies on manual effort. Speech synthesis uses either manual dubbing or text-to-speech (TTS). While TTS boosts efficiency, it struggles with emotional expression, intonation control, and contextual scene adaptation. To address these challenges, we propose DeepDubbing, an end-to-end automated system for multi-participant audiobook production. The system comprises two main components: a Text-to-Timbre (TTT) model and a Context-Aware Instruct-TTS (CA-Instruct-TTS) model. The TTT model generates role-specific timbre embeddings conditioned on text descriptions. The CA-Instruct-TTS model synthesizes expressive speech by analyzing contextual dialogue and incorporating fine-grained emotional instructions. This system enables the automated generation of multi-participant audiobooks with both timbre-matched character voices and emotionally expressive narration, offering a novel solution for audiobook production.

\end{abstract}
\begin{keywords}
Audiobook Synthesis, Text-to-Timbre, Context-Aware Instruct-TTS, Conditional Flow Matching
\end{keywords}

\afterpage{%
  \begin{figure*}[t!]
    \centering
    \includegraphics[width=\textwidth]{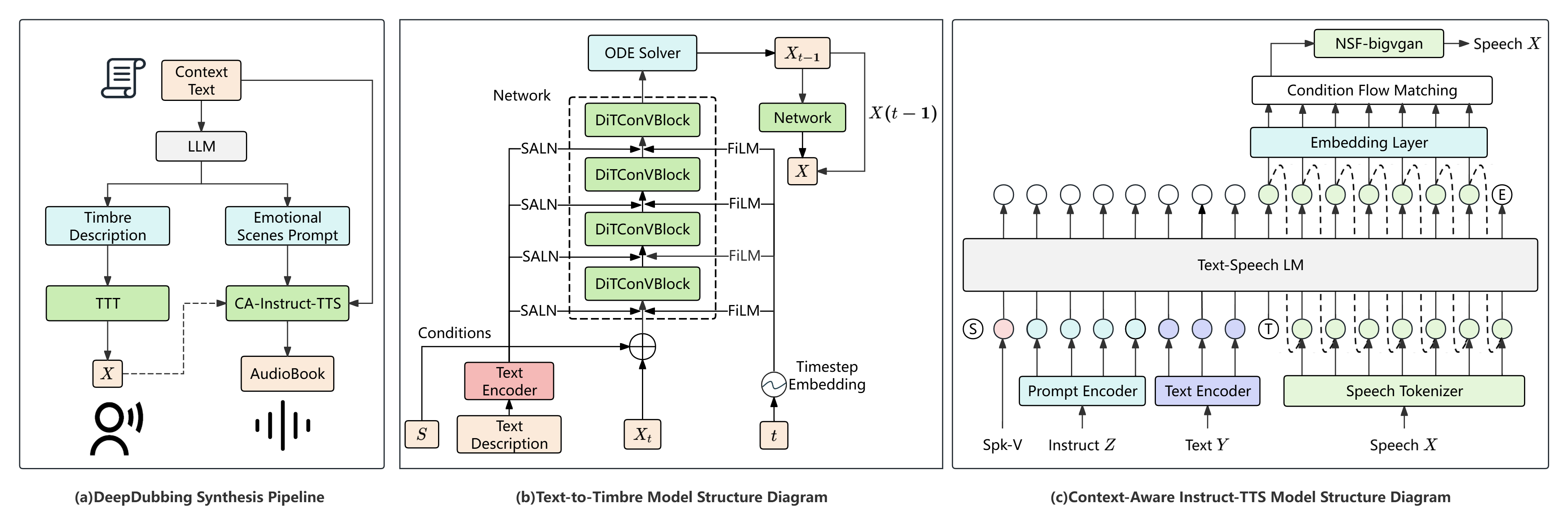}
    \vspace{-2em}
    \caption{An overview of the proposed DeepDubbing. (a) illustrates the end-to-end synthesis pipeline, which consists of two core stages: text-to-timbre generation and emotional scene-based speech synthesis. $X$ denotes the speaker embedding. (b) depicts the model architecture of the Text-to-Timbre module. This model is trained using conditional flow matching, conditioned on gender label $S$ and textual description. (c) shows the overall structure of the CA-Instruct-TTS model, which comprises a text-to-token LLM, a token-to-mel flow matching model, and the NSF-BigVGAN vocoder. The tokens $S$, $E$, and $T$ represent the “start of sequence”, “end of sequence”, and “turn of speech” markers, respectively.}
    \label{fig:model}
    \vspace{-2em}
  \end{figure*}
}

\section{Introduction}
\label{sec:intro}
The multi-participant audiobook segment, which significantly enhances narrative immersion, is experiencing rapid market expansion, generating substantial demand for efficient content generation methodologies.
Traditional production is resource-intensive, as it requires casting multiple voice actors, lengthy recording sessions, and careful direction, all of which result in high costs and long production cycles. Although TTS systems have achieved remarkable naturalness \cite{c25,c26,c27}, automating the generation of high-quality audiobooks with diverse and characteristic voices remains a challenge. This challenge involves two fundamental issues: how to acquire appropriate voice timbres for characters in audiobooks, and how to ensure that the prosody of dialogues, such as emotion, rhythm, and intensity, conforms to the narrative context.

In audiobook voice-timbre analysis, existing methods typically rely on selecting speakers from a predefined timbre list. However, manual selection is costly, and limited voice options fail to meet the demands of hundreds or thousands of audiobook characters. Therefore, we propose a text-to-timbre model to automatically analyze character gender, age, and personality traits to generate suitable voice timbres. Related work on generating timbres from audiobook context remains scarce. DreamVoice \cite{c6} employs a conditional diffusion model to generate speaker embeddings from text prompts, but it struggles to stably coordinate multi-attribute combinations (e.g., age, gender, and personality), resulting in poor consistency. NANSY++ \cite{c23} supports voice design through fine-grained control of continuous attributes (e.g., gender and age) and novel speaker generation. However, it relies on predefined attribute scalars rather than natural language descriptions, thereby limiting its flexibility and interpretability in text-conditioned voice generation.

Furthermore, generating contextually appropriate prosody remains a significant challenge \cite{c1,c4}. Most TTS systems synthesize speech on a sentence-by-sentence basis, lacking broader narrative context---information that voice actors use to determine appropriate emotional delivery \cite{c19,c20,c21}. Without such understanding, synthesized speech often sounds flat and emotionally disconnected, compromising immersion. Recent advances in audiobook synthesis have focused on diverse speaking styles \cite{c3,c5}, context-aware expressiveness \cite{c2}, and long-form prosodic consistency \cite{c4,c20,c21}. TACA-TTS \cite{c2} enhances expressive audiobook synthesis via text-aware and context-aware style modeling, yet lacks fine-grained emotion-scene adaptation for interactive contexts. JELLY \cite{c24} enhances emotion-aware conversational synthesis by leveraging an emotion-text aligned LLM to reason about affective context from dialogue history. However, it lacks explicit modeling of fine-grained scene semantics, limiting its ability to generate speech conditioned on descriptive contextual instructions.

To address these gaps, we propose DeepDubbing, an end-to-end automated system for multi-participant audiobook synthesis. The main contributions of this work include:

\begin{itemize} 
    \item We release BookVoice-50h, a audiobook dataset with annotated timbre profiles and emotion-scene instructions, supporting research on text-to-timbre mapping and expressive TTS.
    \item We propose a conditional flow matching-based Text-to-Timbre (TTT) model that generates accurate speaker embeddings from natural language descriptions, incorporating multi-scale text conditioning and explicit gender control for fine-grained timbre generation.
    \item We develop a Context-Aware Instruct-TTS (CA-Instruct-TTS) system that leverages LLM-derived emotion-scene instructions from narrative context to synthesize expressive speech, effectively mitigating contextual fragmentation.
    \item These components are integrated into an end-to-end pipeline that generates high-quality multi-participant audiobooks directly from text, demonstrating strong potential for industrial-scale applications.
\end{itemize}

\section{Methed}
\label{sec:pagestyle}
The DeepDubbing system consists of two main components: a Text-to-Timbre model that generates speaker embeddings from structured textual descriptions, and a Context-Aware Instruct-TTS model that synthesizes expressive speech conditioned on the generated speaker embedding and context-derived instructions. Details of each component are described in the following subsections.

\subsection{Automated Audiobook Synthesis Pipeline}
We aim to build a fully automated pipeline that converts raw book text into high-quality, multi-speaker audiobooks with context-aware expressiveness.  The overall workflow, depicted in Fig. \ref{fig:model} (a), consists of three main steps:


Step 1: The entire book text is processed by a large language model (LLM), which identifies all characters and generates a structured timbre description for each. This description serves as input to the TTT model to produce a corresponding speaker embedding.


Step 2: The same LLM analyzes the narrative context to generate emotion-scene instructions for each dialogue segment.

Step 3: The CA-Instruct-TTS model synthesizes expressive and contextually appropriate speech for each character based on three inputs: the generated speaker embedding, the current sentence text, and the emotion-scene instruction.

This LLM-powered context parsing and dual-instruction generation mechanism enables fully automated, end-to-end expressive multi-participant audiobook synthesis.

\begin{table*}[t!]
\centering
\caption{Examples of Text-to-Timbre Prompt, Context-Aware Instruct-TTS Prompt.}
\label{tab:examples}
\begin{tabular}{@{}p{\linewidth}@{}}
\hline
\textbf{Text-to-Timbre Prompt}\\
\begin{CJK}{UTF8}{gbsn}  
1. 该角色是一个\textbf{中年男性}，\textbf{身份}是王朝将军，\textbf{性格}铁血威严、霸气侧漏，\textbf{气质}不怒自威
\end{CJK} 
\\
\begin{CJK}{UTF8}{gbsn}  
2. 该角色是一个\textbf{幼儿女性}，\textbf{身份}是世家千金，\textbf{性格}活泼机敏、爱撒娇，\textbf{气质}天真灵动
\end{CJK}
 \\
 \begin{CJK}{UTF8}{gbsn}  
Template: 该角色是一个[幼年、青年、中年、老年][男性、女性]，身份是[xxx]，\textbf{性格}[xxx]
\end{CJK}
 \\
 
\hline
\textbf{Context-Aware Instruct-TTS Prompt} 
\\ 
\begin{CJK}{UTF8}{gbsn}  
1. \textbf{坚定} \textbar \textbf{阵前发布指令时的呐喊} \textbar "三军听令！擂鼓，进军！后退者，斩！"
\end{CJK} 
\\
\begin{CJK}{UTF8}{gbsn}  
2. \textbf{撒娇} \textbar \textbf{向长辈讨要东西时的快语} \textbar "娘亲娘亲！你看那个糖人，翅膀亮晶晶的！给我买一个嘛～"
\end{CJK} 
\\
\begin{CJK}{UTF8}{gbsn}  
Template: [单句情感] \textbar [上下文场景] \textbar [待合成文本]
\end{CJK} \\ 
\hline
\end{tabular}
\vspace{-1em}
\end{table*}

\subsection{Text-to-Timbre Generation via Conditional Flow Matching}
The Text-to-Timbre generation module (architecture shown in Fig. \ref{fig:model} (b)) converts structured textual descriptions into corresponding speaker embedding representations. Built upon a conditional flow matching framework, this module employs a conditional vector field estimator that takes the noised state, text condition, and gender label as inputs to predict the target velocity field. The network adopts a Diffusion Transformer (DiT) \cite{c7} architecture integrated with multi-level fusion mechanisms, specifically Style-Adaptive Layer normalization (SALN) \cite{c13} and  Feature-wise Linear Modulation (FiLM) \cite{c14} to achieve effective condition injection. During training, classifier-free guidance is applied. At inference time, a numerical solver generates high-quality speaker embeddings from random noise that match the textual descriptions, providing reliable input for downstream speech synthesis.

We employ a conditional flow matching framework \cite{c8,c9} for our timbre generation model. In contrast to traditional Diffusion Probabilistic Models (DPM), the adopted Optimal-Transport Conditional Flow Matching (OT-CFM) methodology demonstrates superior performance \cite{c8,c11,c12} through simplified gradient computation, more stable training dynamics, and significantly accelerated sampling speed. Our implementation leverages optimal transport theory to construct a direct probability density path between noise and data distributions, achieving high-quality synthesis with minimal computational overhead. A multi-level conditioning strategy is further incorporated to ensure fine-grained control throughout the generative process.

OT-CFM framework connects a simple prior distribution $x_0 \sim \mathcal{N}(0,1)$ (noise distribution) to the complex data distribution $x_{1} \sim P(x_1)$ through a linear interpolation path, consisting of a forward noising process and a reverse denoising process.

The forward noising process is a fixed, non-learned interpolation procedure. For a target speaker embedding $x_1$, an intermediate state $x_t$ is constructed by linearly mixing it with noise $x_0 $ :
\begin{equation}
\begin{split}
x_t &= (1-(1-\sigma_{min})t) \cdot x_0 + t \cdot x_1
\end{split}
\end{equation}
where the time step $t$ is uniformly sampled from the interval [0,1]. When $t = 0$, $x_t = x_0$ (pure noise); when $t = 1$, $x_t = x_1$ (target data). The corresponding vector field to be regressed is:
\begin{equation}
\begin{split}
u_t &= x_1 - (1 - \sigma_{min})x_0
\end{split}
\end{equation}

The core of OT-CFM is to train a neural network $v_\theta$ to approximate this "direction field" of the forward process, i.e., to estimate the direction from any intermediate point $x_t$ to $x_1$. The vector field estimator $v_{\theta}$ of the TTT model is based on Diffusion Transformer and adopts a multi-level fusion mechanism to integrate conditional information. At the input level, the noised speaker embedding $x_t$ is concatenated with text embedding $c_{\text{t}}$ and gender embedding $c_{\text{s}}$ to form a joint input $[x_t; c_{\text{t}}; c_{\text{s}}]$. For deep feature-wise conditioning, text conditions are injected into each DiT block via SALN, enabling fine-grained control, while timestep information is incorporated using FiLM to guide the denoising trajectory. Our model is trained to regress the true velocity vector $u_t = x_1 - (1 - \sigma_{min})x_0$ using a mean squared error (MSE) loss:

\begin{equation}
\begin{split}
\mathcal{L} = \mathbb{E}_{t,x_{0},x_{1}} \left[ \left\| v_{\theta}(x_{t},t,c_{\text{t}},c_{\text{s}}) - (x_{1}-(1 - \sigma_{min})x_{0}) \right\|^{2} \right]
\end{split}
\end{equation}

During inference, we start from a random Gaussian noise $x_0 \sim \mathcal{N}(0,1)$ and generate a target data sample $x_1$ that conforms to the condition $c$ by solving the ordinary differential equation (ODE) defined by the learned vector field $v_\theta$. We employ an Euler solver for numerical integration:
\begin{equation}
\begin{split}
x_{t+\Delta t}=x_t+v_\theta(x_t,t,c_{\text{t}},c_{\text{s}})\cdot \Delta t
\end{split}
\end{equation}
By integrating from $t = 0$ to $t = 1$, we effectively flow the initial noise to a point in the target data distribution, thus synthesizing a novel speaker embedding that matches the text description $c_t$.

\afterpage{%
\begin{table*}[t!]
  \centering
  \caption{Comparison of TTT performance across age groups: Sex Accuracy (SA), Age Accuracy (AA) and Character Matching Score (CMS)  with 95\% CI ( CMS Score: 0=Irrelevant, 1=Marginally Relevant, 2=Partially Consistent, 3=Highly Consistent, 4=Excellent Match).}
  \label{tab:sa_aa_age}
  \renewcommand{\arraystretch}{1.15}
  \setlength{\tabcolsep}{2.4pt}
  \begin{tabular*}{\textwidth}{@{\extracolsep{\fill}}c|cccc|cccc|c@{}}
    \hline
    & \multicolumn{4}{c|}{SA (\%) $\uparrow$}
    & \multicolumn{4}{c|}{AA (\%) $\uparrow$}
    & \\
    \cline{2-9}
    \multicolumn{1}{c|}{\raisebox{1ex}{\textbf{Method}}}
    & \textbf{Child} & \textbf{Youth} & \textbf{Middle} & \textbf{Elder}
    & \textbf{Child} & \textbf{Youth} & \textbf{Middle} & \textbf{Elder}
    & \multicolumn{1}{c}{\raisebox{1ex}{\textbf{CMS} $\uparrow$}} \\ 
    \hline
    TTT-T5-Large
    & 90.00 & 98.75 & 99.38 & 98.75
    & 23.13 & \textbf{77.50} & 57.50 & 46.88
    & 2.375$\pm$0.038\\
    TTT-Roberta-Large
    & \textbf{98.13} & 95.63 & \textbf{100.00} & \textbf{100.00}
    &  16.25  & \textbf{77.50} & 75.63 & 69.38
    & 2.359$\pm$0.044\\
    TTT-Qwen3-0.6B
    & 96.25 & \textbf{100.00} & \textbf{100.00} & \textbf{100.00}
    & \textbf{74.38} & 74.38 & \textbf{90.00} & \textbf{73.13}
    & \textbf{2.866$\pm$0.036}\\
    \hline
  \end{tabular*}

\end{table*}
}

\subsection{Context-Aware Instruct-TTS Synthesis}
As shown in Fig. \ref{fig:model} (c), our CA-Instruct-TTS model adopts a structure inspired by CosyVoice \cite{c25,c26}, comprises three core components: a large language model (LLM), a conditional flow matching model, and a vocoder. 
The input to the LLM is formed by concatenating four components along the token dimension:
\begin{equation}
\begin{split}
Input = E_{spk}  \oplus T_{instruct} \oplus T_{text} \oplus T_{speech}
\end{split}
\end{equation}
where $E_{spk}$ denotes the speaker embedding vector, $T_{instruct}$ represents the subword tokenization of the natural language instruction, $T_{text}$ is the subword sequence of the target text, 
$T_{speech}$ indicates the acoustic unit sequence produced by the speech tokenizer, and $\oplus$ denotes the concatenation operation along the token dimension. The language model component is continuously trained from a QinYu-based speech model (an internal, closed-source base model), employing a 12-layer Transformer architecture to learn the mapping from multimodal conditions to acoustic features.

In contrast to CosyVoice, we introduce several architectural refinements. On the flow-matching side, we employ a DiT network to solve the flow-matching distribution mapping problem, with a structure that remains consistent between the CA-Instruct-TTS and TTT model. For the vocoder, we replace the original design with an NSF-BigVGAN model \cite{c15,c16} to enhance audio quality. Together, these modifications enable high-fidelity, expressive speech synthesis.

\section{EXPERIMENTS}
\label{sec:typestyle}
\subsection{Experimental Settings}
\subsubsection{Dataset}
We employ a large-scale internal multi-participant audiobook dataset comprising over 4,000 hours of high-quality speech. An automated LLM-based annotation pipeline generates structured labels for two tasks. For the Text-to-Timbre task, it constructs over 300K text-based timbre descriptions following a Gender\textbar Age\textbar Personality \textbar Identity template; for the Context-Aware Instruct-TTS task, more than 2 million instructions are generated under an Emotion\textbar Contextual Scenario template, covering over 44 fine-grained emotion categories. To support both tasks, the Cam++ model \cite{c22} is employed to extract speaker embeddings for each speech segment during training. The test set contains only unseen speaker identities to facilitate evaluation.

To advance research in text-guided voice generation and expressive audiobook synthesis, we release BookVoice-50h, a novel synthetic dataset generated by our proposed models to support two tasks: TTT and CA-Instruct-TTS. The templates for text descriptions of the TTT model and instructions for CA-Instruct-TTS are shown in Table \ref{tab:examples}. The BookVoice-50h synthetic dataset will be released on Hugging Face upon publication. For audio samples and generation capabilities, visit our demo page: https://tme-lyra-lab.github.io/DeepDubbing.

\subsubsection{Model Architecture and Training Configuration}
The TTT model employs a conditional flow matching architecture with a 4-layer DiT backbone comprising 4 attention heads and 392 hidden dimensions. Text conditions are encoded using Qwen3-Embedding-0.6B, projected to 192 dimensions, and injected into the network via SALN, while timestep and gender labels are incorporated through FiLM modulation and concatenation, respectively. Classifier-free guidance (CFG) \cite{c10} is applied with a conditional dropout rate of 0.2; during inference, the CFG scale and rescale factor are set to 3.0 and 0.7. 

The CA-Instruct-TTS model utilizes a 12-layer Transformer language model that takes four inputs: a timbre embedding, a text sequence, a context instruction, and an acoustic unit sequence. The flow-matching component is conditioned on the timbre embedding, speech token sequence, and masked acoustic features, and an NSF-BigVGAN vocoder is adopted for waveform synthesis. The language model is continuously trained based on the QinYu-based speech model.

\subsection{Evaluation Metrics}
To comprehensively evaluate the performance of the proposed system, we employ both subjective and objective metrics assessing intelligibility, naturalness, and attribute consistency for both TTT and CA-Instruct-TTS modules. All subjective tests were carried out by rigorously screened and trained expert listeners.

The TTT module was evaluated by synthesizing audio waveforms from generated speaker embeddings using the CA-Instruct-TTS module. The test set comprises 80 samples balanced across age, gender, and personality traits. Assessment includes Character Matching Score for character trait matching (CMS, 4-point scale), along with gender and age accuracy.

The CA-Instruct-TTS module was evaluated for expressiveness and speech quality. The model synthesized 195 utterances covering over 44 emotion categories with balanced distribution. The evaluation comprises Mean Opinion Score for emotion (MOS-E) and naturalness (MOS-N), along with Word Error Rate (WER) computed using the Whisper-large-v3 model \cite{c17}.

\begin{table}[t!]
\centering
\caption{Comparison of subjective and objective scores by CA-Instruct-TTS: Word Error Rate (WER), and Mean Opinion Scores (MOS):  N-Naturalness, E-Emotion.}
\label{tab:model_performance}
\renewcommand{\arraystretch}{1.15}
\setlength{\tabcolsep}{4pt}
\begin{tabular*}{\linewidth}{@{\extracolsep{\fill}}l|c|cc@{}} 
\hline
\textbf{Method} &
\textbf{WER}\,$\downarrow$ &
\textbf{MOS-N}\,$\uparrow$ & \textbf{MOS-E}\,$\uparrow$ \\
\hline
CA-TTS        & \textbf{2.39\%}  & 3.10 & 3.67 \\
\hline
CA-Instruct-TTS & 2.54\%  & \textbf{3.33} & \textbf{4.15} \\
\hline
\end{tabular*}
\end{table}

\begin{figure}[t]
  \centering
  \includegraphics[width=\linewidth]{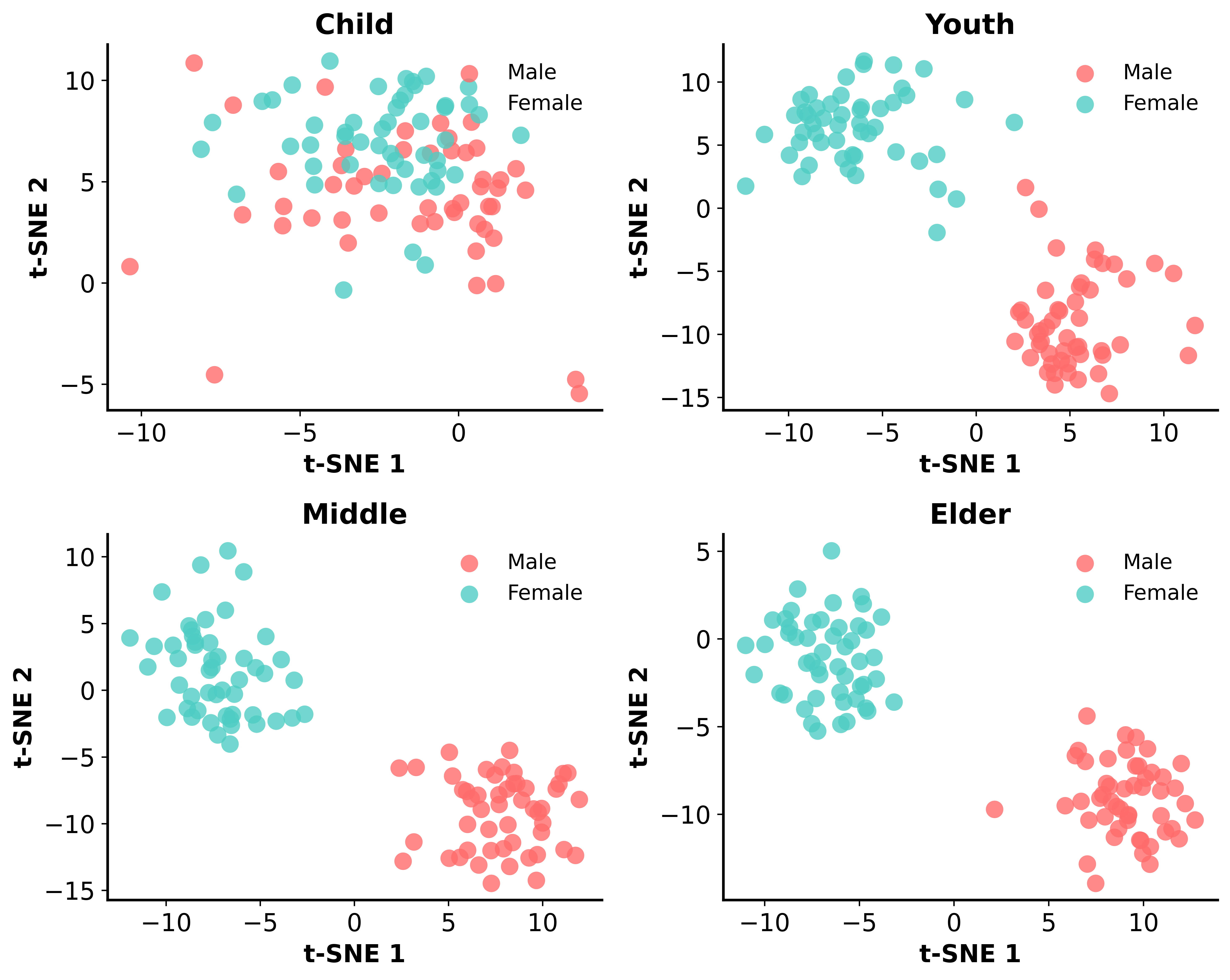}
  \caption{t-SNE Analysis of Gender Clustering in Age-Stratified Speaker Embeddings.}
  \label{fig:single}
\end{figure}

\subsection{Experimental Results}
We conducted comprehensive evaluations of our proposed system. As shown in Table \ref{tab:sa_aa_age}, we compared three text encoders for the TTT module. The results demonstrate that TTT-Qwen3-0.6B, compared to TTT-T5-Large, and TTT-Roberta-Large, achieves the best performance or is highly competitive across all metrics, validating the effectiveness of Qwen3-Embedding-0.6B in understanding and generating complex timbre semantic descriptions. As illustrated in Fig. \ref{fig:single}, gender classification accuracy for children is significantly lower. This observation is consistent with findings in NANSY++ \cite{c23}, which reports that pre-pubertal children's voices exhibit high acoustic similarity and show poorly differentiated gender characteristics, posing challenges for embedding-based discrimination. Furthermore, the presence of adult female speech imitating child voices in existing child speech datasets further confounds gender differentiation.

For speech synthesis quality evaluation, we compared CA-Instruct-TTS with a baseline approach that directly inputs text and speech to the LLM without instruction guidance, as presented in Table \ref{tab:model_performance}. The experimental results demonstrate that CA-Instruct-TTS achieves the best performance or is highly competitive across all subjective and objective metrics. While maintaining comparable word error rates, our proposed method demonstrates significant improvements in both naturalness (MOS-N) and emotional expressiveness (MOS-E). The enhanced emotional expressiveness particularly highlights the effectiveness of our context-aware instruction mechanism in generating more expressive and contextually appropriate speech synthesis. The improved naturalness scores indicate that the instruction-based approach better captures the intended emotional and contextual nuances, thereby producing more human-like speech generation.

In summary, both core modules of the DeepDubbing system exhibit exceptional performance in their respective tasks, demonstrating the effectiveness and advancement of the proposed approach in automated audiobook synthesis.

\section{CONCLUSION AND FUTURE WORK}
\label{sec:majhead}
This paper presents DeepDubbing, an automated speech synthesis system designed for multi-participant audiobook generation. 
Experimental results demonstrate that our system achieves strong performance in speech naturalness, timbre matching accuracy, and emotional expressiveness. However, the TTT model shows limitations in generating child-like voices, primarily due to the scarcity of genuine child speech samples in the training data---most existing data consists of adult-imitated child voices. Future work will prioritize three key directions: enhancing youth voice generation through the collection of more authentic child speech data, developing finer-grained timbre control mechanisms to enable more precise manipulation of vocal characteristics, and extending the system to multilingual scenarios to broaden its practical applicability.

\clearpage              




\bibliographystyle{IEEEbib} 
\bibliography{Template_new}

\end{CJK}
\end{document}